\documentstyle[12pt,twoside]{article}
\input{amssym.def}
\input{amssym}
\newtheorem{proposition}{Proposition}
\newtheorem{lemma}{Lemma}
\newtheorem{corollary}{Corollary}

\def\mod{{\rm mod}}

\def\ha{Ha\-mil\-to\-nian}

\def\lsj{\left(\frac{j}{p}\right)}

\def\lsxi{\left(\frac{x}{p_i}\right)}

\def\R{\Bbb R}
\def\Z{\Bbb Z}
\def\T{\Bbb T}
\def\N{\Bbb N}

\def\C{\Bbb C}
\def\la{\langle}
\def\be{\begin{equation}}
\def\ee{\end{equation}}
\def\ra{\rangle}
\def\ds{\displaystyle}
\begin{document}
\date{}
%
%
%
%
\title{Ground states for a class of deterministic spin models with
glassy
behaviour}

\author{
I.Borsari, S.Graffi and F.Unguendoli
 \\Dipartimento di Matematica, Universit\`{a} di Bologna\\40127 Bologna,
Italia}
\maketitle
\begin{abstract}
 {We consider the deterministic model with glassy behaviour, recently
introduced by Marinari, Parisi and Ritort, with \ha\
$H=\sum_{i,j=1}^N J_{i,j}\sigma_i\sigma_j$, where $J$ is the discrete
sine
Fourier transform.
The ground state found by these authors for $N$ odd and $2N+1$ prime is
shown to
become asymptotically dege\-ne\-ra\-te when $2N+1$ is a product of odd
primes, and to
disappear for $N$ even. This last result is based on the explicit
construction
of a set of eigenvectors for
$J$, obtained through its formal identity with the
imaginary part of the propagator of the quantized unit symplectic matrix
over
the $2$-torus.}
\end{abstract}
\baselineskip = 18pt
\section{Introduction}
\setcounter{equation}{0}%
\setcounter{theorem}{0}%
\setcounter{proposition}{0}%
\setcounter{lemma}{0}%
\setcounter{corollary}{0}%
\setcounter{definition}{0}%

It has been recently established \cite{1},\cite{2},\cite{3} that a wide
class of deterministic,
infinite-range deterministic Ising spin models does actually exhibit the
glassy
behaviour of the random coupling case, with the important difference,
however,
that the mean field equations of the model, derived by Parisi and
Potters\cite{4}(hereafter the PP equations), are not the standard TAP
equations.
 Unlike the random case(see e.g.[5]), they do not determine the critical
temperature of the glassy transition by linearization around the largest
eigenvalue of the interaction matrix. \par Among these models
a special role is played by the so-called {\it sine} (or, equivalently,
{\it
cosine}) model, in which the interaction matrix ${\ds J=(J)_{i,j},
i,j=1,\ldots,N; N\in\N}$ among $N$ spins (with periodic boundary
conditions) defining the \ha
\be
\label{Ha}
H=-\frac12\sum_{i,j=1}^N \,J_{i,j}\sigma_i\sigma_j
\ee
is given by
\be
\label{J}
J_{i,j} =\frac{2}{\sqrt{2N+1}}\;{\rm sin}\left(\frac{2\pi
ij}{2N+1}\right),\quad i,j=1,\ldots,N
\ee
namely by twice the uppermost left block of
the the discrete sine (cosine) Fourier transform
\be
\label{S} S_{i,j} =\frac{1}{\sqrt{2N+1}}\;{\rm sin}\left(\frac{2\pi
ij}{2N+1}\right), \quad i,j=1,\ldots,2N
\ee
The factor $2$ accounts for the orthogonality of $J$. Here $N$ is odd
and
$p=2N+1$ prime, i.e.
$p$ is a prime of the form $p=4m+3$. In fact, in this case the ground
state
configuration can be explicitly computed\cite{3} and is given by
${\ds \sigma_j=\left(\frac{j}{p}\right)\;j=1,\ldots,N}$. Here ${\ds
\left(\frac{j}{p}\right)}$ is the Legendre symbol of $j$, namely
\be
\label{Leg}
\left(\frac{j}{p}\right)=\left\{\begin{array}{l} +1, \;{\rm if}\;
j\equiv x^2\;(\mod \; p) \\ -1, \;{\rm if}\;
j\not\equiv x^2\;(\mod \; p) \end{array}\right.
\ee
where $x\in (0,1,\ldots,p-1)\equiv \Z_p=\Z(\mod\;p)$. In other words, if
$j$
is a quadratic residue of $p$ its Legendre symbol is $1$, and $-1$ in
the
opposite case. The existence of such a complex ground state, proved by
showing that on the spin configuration defined by the Legendre symbols
the
energy actually assumes its absolute minimum ${\ds -\frac{N}{2}}$,
yields the
possibility of numerically detecting a first-order "crystalline" phase
transition at a temperature higher than the critical one for the glassy
transition\cite{3,4} and of explicitly finding\cite{4} the corresponding
solution of the PP mean field equations under the form ${\ds
m_i=\sqrt{q}\left(\frac{i}{p}\right)}$. Here as usual
$m_i$ denotes the magnetization on the site $i$, and ${\ds
q=\frac1N\sum_{i=1}^N m_i^2}$ the Edwards-Anderson order parameter.  The
reader is referred to\cite{4}, \S3 for a discussion of the relevance of
the
existence of crystalline phase on the glassy behaviour of the
system.\par  The existence
of such a ground state critically depends on the arithmetic restrictions
on
$N$ (actually Parisi and Potters\cite{4} give arguments supporting its
disappearance for
general values of $N$) and hence it can be of interest to look into the
question from a
rigorous point of view.
\par
 We can distinguish two cases for $p$ odd:
\begin{itemize}
\item[(A)] $p$ is of the form $4m+3$ (the case considered by Marinari,
Parisi and Ritort when $p$ is prime);
\item[(B)] $p$ is of the form $4m+1$.
\end{itemize}
In case (A) we show that, when the restriction on the primality of
$p$ is essentially removed, namely when
$p=2N+1=p_1p_2\ldots p_s$ is the product of
$s$ distinct primes such that $2N+1=4m+3$ (the factorization of $p=2N+1$
consists in an odd number
$t$ of primes $p_i$ of the form $4m_i+3$ and of an arbitrary number of
primes
of the form $p_i=4m_i+1$), then
\begin{itemize}
\item[(1)] The ground state energy ${\ds -\frac{N}{2}}$ be\-co\-mes
a\-sym\-pto\-tical\-ly
degenerate of or\-der \linebreak ${\ds D=O(2^{N^{\frac{s-1}{s}}})}$,
namely there are $D$
distinct spin configurations $\sigma_j^{(l)}: j=1,\ldots,p$,
$l=1,\ldots,D$
such that their energy $E(\sigma_j^{(l)})$ fulfills the estimate
\be
E(\sigma_j^{(l)}) \leq -\frac{N}{2}\left(1-KN^{-\frac{1}{s}}\right)
\ee
for some positive constant $K$ independent of $l$.
\item[(2)]
The $D$ distinct spin configurations $\sigma_j^{(l)}$ are obtained as
follows
\be
\label{q_m}
 \sigma_j^{(l)}=\left\{\begin{array}{ll} \psi(j),& \quad {\rm if}\quad
\psi(j)\neq 0 \\
\pm 1 & \quad {\rm if}\quad \psi(j)= 0
\end{array}\right.
\ee
where $\psi(j)$ is the Jacobi symbol of $j\in\Z_p$ with respect to
$p=2N+1$:
\be
\label{Jacobi}
\psi(j)=\prod_{i=1}^s\left(\frac{j}{p_i}\right), \qquad
\left(\frac{j}{p_i}\right)=0 \quad {\rm if}\quad (j,p_i)=0
\ee
(here $(k,p)$ denotes the MCD between $k$ and $p$; $(k,p)=0$ means that
$k$ is a
multiple of $p$) and the number of zeroes of
$\psi(j)$ behaves like ${\ds N^{\frac{s-1}{s}}}$ for $N$ large;
\item[(3)]
For $q$ suitably small, the magnetization vectors $ m_j^{(l)}=
\sqrt{q} \sigma_j^{(l)}$ solve the PP mean field equations
\be
\label{PP}
m_i={\rm tanh}\left(2\beta G^{\prime}(\beta(1-q))m_i-\beta\sum_{j=1}^p
J_{ij}m_j\right)
\ee
where ${\ds G(x)=-\frac14{\rm ln}\left(\frac{1+\sqrt{1+4x^2}}{2}\right)+
\frac14\left(\sqrt{1+4x^2}-1\right)}$.
\end{itemize}
We will recall later how the above properties allow us an immediate
application of the argument of Parisi and Potters (\cite{4}, \S4) to
strongly
support the conclusion that in this case there are $D$ "crystal" states
with
properties analogous to the one existing for $p=4m+3$ prime. \par In
case (B)
we consider the case analogous to that of \cite{3}, namely $p=2N+1$
prime
of the form $4m+1$, and show that $J$
cannot admit eigenvectors whose components are all of the form
$\pm 1$ in correspondence to the eigenvalues $\pm 1$. Hence the minimum
of
the energy quadratic form $\la u, Ju\ra$ is never reached when $u$ is a
spin
configuration, let alone the configuration of the Legendre symbol valid
for $p=4m+3$. This represents (for such values of $N$) a rigorous proof
of the
conjecture of \cite{4}, and hence suggests disappearance of the above
"crystal state"
picture.
\par In the forthcoming Sect.2 we describe the number theoretic argument
proving Properties
(1),(2) and (3) of case (A), and in Sect.3 we show (B), basing the proof
on the explicit
construction of a set of eigenvectors for the operator $J$.  \par This
construction can
be interesting in itself because it is based on the metaplectic
representation of the
quantized symplectic linear maps over the $2-$torus. In particular, the
operator $S$
turns out to coincide\cite{10},\cite{11} with (the imaginary part of)
the operator
quantizing of the standard unit symplectic
$2\times 2$ matrix
\[
I_{sp}= \left(\begin{array}{cc} 0 & -1 \\ 1 & 0\end{array}\right)
\]
This operator is a $N\times N$ unitary matrix, $N$ being the inverse of
the
Planck constant, so that in this context
the thermodynamic limit $N\to\infty$ is formally equivalent
to the classical limit.

\vskip 1.0cm
\section{A real eigenvector of the operator $J$}
\setcounter{equation}{0}%
\setcounter{theorem}{0}%
\setcounter{proposition}{0}%
\setcounter{lemma}{0}%
\setcounter{corollary}{0}%
\setcounter{definition}{0}%
Marinari, Parisi and Ritort\cite{3}
prove that for $p=4m+3=2N+1$ the ground state is given by the spin
configuration
defined by the Legendre symbols
\[
 \sigma_L=(\sigma_1,\ldots,\sigma_N);\qquad \sigma_j=\lsj, j=1,\ldots,N
\]
by explicit verification that $\sigma_L$ is an absolute minimizer of the
energy. As a consequence, $\sigma_L$ must necessarily be an eigenvector
of the
operator
$J$ defined by (\ref{J}) corresponding to the eigenvalue $1$
(recall that $J$ is a real orthogonal matrix so that its spectrum
consists only of the eigenvalues $\pm 1$). In analogy with this result,
in
the present case the basic step is represented by the construction of an
eigenvector of the operator $J$  whose components are all
$\pm 1$ or $0$ because it is defined by the Jacobi symbol ${\ds
\chi_J(x)=\prod_{i=1}^s\lsxi, x=1,\ldots,p}$. \par
The construction of this
eigenvector will be an easy consequence of the following
\begin{lemma}
\label{L}
Let ${\ds N=\prod_{i=1}^s p_i}$ be the product of $s$ pairwise different
odd
primes such that $N=4m+3$. Then the matrix $S$ defined by (\ref{S}),
whose elements are
\be
\label{2.1}
S_{k,x}=\frac1{\sqrt{N}}{\rm \sin}\left(\frac{2\pi}{N}kx\right)
\ee
admits the vector ${\ds
\chi_J(x)=\prod_{i=1}^s\lsxi, x=1,\ldots,N}$ as an eigenvector
corresponding to
the eigenvalue $1$.
 \end{lemma}
{\it Proof.} It is a classical result in number theory (see
e.g.\cite{7}, Proposition A.7) that $\chi_J(x), x\in\Z_N$ is the unique
primitive multiplicative character of the ring $\Z_N=\Z(\mod N)$, and it
is
also well known (see again \cite{7}, Proposition 2.1 or \cite{8},
Theorem 8.15)
that the Gaussian sum
\be
\label{2.2}
\tau_k(x)=\sum_{k=1}^N\chi(x)e^{\frac{2\pi i}{N}kx}
\ee
is separable for all $k$ if $\chi$ is a primitive multiplicative
character.
Namely, one has
\be
\label{2.3}
\tau_k(\chi)=\sum_{k=1}^N\overline{\chi}(k)\tau_1(\chi)
\ee
On the other hand, Theorem 2.1 of \cite{7} states that, if $\chi$ is any
real primitive
character
\be
\label{2.4}
\tau_1(\chi)=\left\{\begin{array}{l} N^{\frac12} \quad {\rm if}\quad
\chi(-1)=1\\ iN^{\frac12} \quad {\rm if}\quad
\chi(-1)=-1 \end{array}\right.
 \ee
Therefore we get in our case
\begin{eqnarray*}
\label{2.5}
(S\chi_J)_k=\frac1{\sqrt{N}}\sum_{x=1}^{N}{\rm
\sin}\left(\frac{2\pi}{N}kx\right)\chi_J(x)
=\frac1{2i\sqrt{N}}\left(\chi_J(k)-\chi_J(-k)\right)\sum_{x=1}^N\chi_J(x)
e^{\frac{2\pi i}{N}kx}\\
=\frac1{2i\sqrt{N}}\left(\chi_J(k)-\chi_J(-k)\right)i\sqrt{N}
=\frac12\left(\chi_J(k)-\chi_J(-k)\right)
 \end{eqnarray*}
since $\chi_J(-1)=-1$ if $N=4m+3$ (see e.g.\cite{8}, Theorem 9.10). Now,
if
$(k,p_i)>1$ for at least one $i$, then
$\chi_J(-k)=0=\chi_J(k)$ by definition. Let now $(k,p_i)=1$ for all
$1\leq i \leq s$. By
the multiplicative property of the Legendre symbols we have
\[
\left(\frac{-k}{p_i}\right)=
\left(\frac{-1}{p_i}\right)\cdot\left(\frac{k}{p_i}\right)=
-\left(\frac{k}{p_i}\right) \qquad \left(\frac{-k}{p_j}\right)=
\left(\frac{-1}{p_j}\right)\cdot\left(\frac{k}{p_i}\right)=
\left(\frac{k}{p_i}\right)
\]
because $-1$ is quadratic residue of all primes of the form $4m+1$ and
non
residue of the primes of the form $4m+3$ (\cite{8}, Theorem 9.10). Hence
we can write
\begin{eqnarray*}
\label{2.6}
\chi_J(-k)=  \prod_{\stackrel{i=1}
{p_i=4m_i+3}}^t\left(\frac{-k}{p_i}\right)\cdot
\prod_{\stackrel{j=1}
{p_j=4m_j+1}}^{s-t}\left(\frac{-k}{p_j}\right)  \\
= (-1)^t\prod_{\stackrel{i=1}
{p_i=4m_i+3}}^t\left(\frac{k}{p_i}\right)\cdot
\prod_{\stackrel{j=1}
{p_j=4m_j+1}}^{s-t}\left(\frac{k}{p_j}\right) =-\chi_J(k)
  \end{eqnarray*}
Therefore
\[
(S\chi_J)(k)=\frac12(\chi_J(k)-\chi_J(-k))=\chi_J(k)
\]
and this concludes the proof of the Lemma. \vskip 0.2cm
We proceed now to the construction of the real eigenvector of $J$
with components $(\pm 1,0)$.
\begin{corollary}
\label{C1}
Let $N$ be such that ${\ds p\equiv 2N+1=\prod_{i=1}^sp_i}$ where
$p_i:i=1,\ldots,s$
are $s$ distinct primes
such that there is an odd number $t$ of primes $p_i$ of the form
$4m_i+3$. Let
$\psi\in\R^n$ be the vector formed by the first $N$ components of the
real
primitive character ${\ds \chi_J(x) (\mod\; p=2N+1)}$ defined in Lemma
\ref{L}
above.
\par\noindent
Then the operator $J$ acting on $C^N$, defined by the matrix
(\ref{J}), admits
$\psi$ as an eigenvector corresponding to the eigenvalue
$1$.
\end{corollary}
{\it Proof.} We have
\[
J_{k,x} =\frac{2}{\sqrt{2N+1}}{\rm sin}\left(\frac{2\pi kx}{2N+1}\right)
\]
Now the matrix $S$ is row antisymmetric, and the $2N+1$-th row vanishes.
We have
seen above that $\chi_J(-k)=-\chi_J(k)$. Then, by the former Lemma
\begin{eqnarray*}
(J\psi)_k =
\frac{2}{\sqrt{2N+1}}\sum_{x=1}^N{\rm sin}\left(\frac{2\pi
kx}{2N+1}\right)\psi(k) \\ =\frac{2}{2\sqrt{2N+1}}\sum_{x=1}^{2N+1}{\rm
sin}\left(\frac{2\pi kx}{2N+1}\right)\psi(k)=\chi_J(k)
\end{eqnarray*}
where now $k=1,\ldots,N$. This proves the corollary.\vskip 0.2cm
Let us now turn to the proof of Assertions (1), (2), (3) stated in the
Introduction. Consider first the simplest possible case, given by
$N=p_1p_2$,
where $p_1$ is of the form $4m+3$ and $p_2$ of the form $4m+1$, so that
$N$
is of the form $4m+3$. We can assume (see e.g.\cite{7}) that $|p_1-p_2|$
is independent of $N$, so that $p_1\sim \sqrt(N), p_2\sim \sqrt(N)$ as
$N\to\infty$. With $p=2N+1$, consider the eigenvector $\psi(x):
\psi(x)\in\{-1,0,1\}$ of the above corollary. Remark that the zero
components
of the eigenvector are obtained in correspondence of the multiples of
$p_1$ or
$p_2$ between $1$ and $N$. There are at most
\[
h=\frac{p_1-1}{2}+\frac{p_2-1}{2}
\]
such multiples, and, since $p_{1,2}\sim \sqrt{N}$, we have $h\sim
\sqrt{N}$.
Hence the energy $E(\psi)$ of the vector $\psi$ (note that this vector
is not a
spin configuration) is given by
\begin{eqnarray*}
\label{2.7}
E(\psi)=-\frac12 \sum_{i,j=1}^N S_{i,j}\psi_i\psi_j=
-\frac12\left(|\psi^{+}|^2-|\psi^{-}|^2\right) \\
\sim
-\frac12(N-\sqrt{N})=-\frac{N}{2}\left(1-\frac{\sqrt{2}}{\sqrt{N}}\right)
\end{eqnarray*}
Here $\psi^{+}$ and $\psi^{-}$ denote the projection of $\psi$ on the
eigenspaces $V^{\pm}$ corresponding to the eigenvalues $1$ and $-1$,
respectively.
\\ Now out of $\psi$ we can define ${\ds D=2^{h}}$ spin configurations
in the following way
\be
\label{2.8}
\sigma_x=\left\{\begin{array}{l} \psi(x) \quad {\rm if}\quad
\psi(x)\neq 0 \\
\pm 1 \qquad {\rm if}\quad  \psi(x)= 0\end{array} \right. \quad
x=1,\ldots,N
\ee
Set now $v=\sigma-\psi$. The energy ${\ds
E(\sigma)=-\frac12\left(\|\sigma^{+}\|^2-\|\sigma^{-}\|^2\right)}$ is
obviously
maximal when $v\in V^{-}$. Therefore, since $v$ has at most $\sqrt{N}$
non-zero components and $\psi$ is an eigenvector corresponding to the
eigenvalue $1$ of $S$, we have
\begin{eqnarray*}
E(\sigma)\leq -\frac12\left(\|\psi\|^2-\|v\|^2\right) \\ \sim
-\frac12(N-\sqrt{N})=-\frac{N}{2}\left(1-\frac1{\sqrt{N}}\right)
\end{eqnarray*}
Hence the energy of all $D$ states $\psi^l:l=1,\ldots,D$ tend to the
minimum energy
${\ds -\frac{N}{2}}$ as $N\to\infty$. \par
There is now no difficulty in extending the argument to the general case
stated
in \S1, in which $N=p_1\times\ldots\times p_s$ with $N=4m+3$ and
$p_1<p_2<\ldots<p_s$ odd primes. We assume that there is a constant $C$
(depending on $s$) such that $p_s\leq Cp_1$. By repeating the above
argument
one easily obtains that in this case the number $h$ of the zero
components of
the eigenvector $\psi$ of $S$ fulfills the estimate
\be
\label{2.9}
h\sim A N^{\frac{s-1}{s}}
\ee
for some constant $A$ indepedent of $N$ (but dependent on $s$). Hence,
as
above, we can construct ${\ds D=2^{h}}$ spin configurations
$\sigma$ whose energy $E(\sigma)$ fulfills the estimate
\be
\label{2.10}
E(\sigma) \leq -\frac{N}{2}\left(1-\frac{K}{N^{1/s}}\right)
\ee
for some $K$ independent of $s$, and thus the ground state is
asymptotically
degenerate of order $D$ as $N\to\infty$. This concludes the verification
of
Assertions (1) and (2) of \S1.
\vskip 0.2cm
The verification of Assertion (3) proceeds exactly as in \cite{4}: the
ansatz
\[
m_i=\sqrt{q}\epsilon_i, \qquad q=\frac1{N}\sum_{i=1}^N m_i^2
\]
where the $\{\epsilon_i\}$ are $\pm 1$ or $0$ reduces the PP equations
\ref{PP}
to
\be
\label{PPP}
q={\rm
tanh}^2\left\{\beta\sqrt{q}\left[1+\frac{1-\sqrt{1+4\beta^2(1-q)^2}}
{2\beta(1-q)}\right]\right\}
\ee
Since we can take for
$\{\epsilon_i\}$ any one of the eigenvectors $\psi^l;\; l=1,\ldots,D$ of
$S$
constructed before, we see that the magnetization vectors of
components $m_i^l=\sqrt{q}\psi_i^l$ yield $D$ solutions of the mean
field
equations (\ref{PP}) provided $q$ solves (\ref{PPP}). Now (\ref{PPP})
always
admits the paramagnetic solution $q=0$ and hence, as in \cite{4}, for
$\beta$ large enough will also admit a solution for $q\neq 0$. Moreover,
the
specific Gibbs free energy $\beta f_l$ of all solutions will be given by
(26)
of \cite{4} up to an error of order ${\ds N^{-\frac{1}{s}}}$, namely
\be
\label{EL}
\beta f_l=\frac{1+\sqrt{q}}{2}\;{\rm
ln}\left[\frac12(1+\sqrt{q})\right]+
\frac{1-\sqrt{q}}{2}\;{\rm ln}\left[\frac12(1-\sqrt{q})\right]
-\frac{\beta}{2}q-G(\beta(1-q))+O(N^{-\frac{1}{s}})
\ee
In fact, the total Gibbs free energy $\beta\Phi$ as a function of the
magnetizations $m_i$ is given by formula (19) of
\cite{4}
\begin{eqnarray*}
\beta\Phi =  \frac12\sum_{i=1}^N\left\{(1+m_i){\rm
ln}\left[\frac12(1+m_i)\right]+(1-m_i){\rm
ln}\left[\frac12(1-m_i)\right]\right\} \\
 -\frac{\beta}{2}\sum_{i,j=1}^NS_{i,j}m_im_j -NG(\beta(1-q)
\end{eqnarray*}
Taking $m_i=\sqrt{q}\psi^l_i$ we get (\ref{EL}) because ${\ds
\sum_{i,j=1}^NS_{i,j}\psi^l_i\psi^l_j=q+O(N^{-\frac{1}{s}})}$. \par
Therefore we can apply directly the results of the numerical analysis
of\cite{4} showing that (\ref{PPP}) admits a solution with $q=0.92$ for
$T<0.400$ to conclude that there $D$ solutions with such $q$, which for
$N$ large enough will have free energy bigger than that of the
parametric
solutions as long as $T>0.178$, and smaller for $T<0.178$ so that the
absolute minimum of the specific free energy is also asymptotically
degenerate.  Therefore we can conclude that the picture of the "crystal"
state should persist also in this situation, up to a degeneracy.
\vskip 1.0cm\noindent
\section{The case of $p$ prime of the form $4m+1$. Explicit
construction of the eigenvectors of $S$}
 Let us first proceed to the construction of a set
of eigenvectors for
$S$. We start from the obvious constatation that this operator is the
imaginary
part of the discrete Fourier transform, defined as
\be
\label{F}
{\cal F}\psi_l=\frac1{\sqrt{p}}\sum_{k=0}^{p-1}e^{\frac{2\pi
i}{p}}\psi_k
\ee
Namely,
\be
\label{F1}
S=\frac{{\cal F}-{\cal F}^{-1}}{2i}=\frac{{\cal F}-{\cal F}^{\ast}}{2i}
\ee
The discrete Fourier transform operator
$\cal{F}$ coincides with the unitary  evolution operator $V_J$
quantizing,
via canonical (see \cite{10}) or, equivalently, geometric (see
\cite{11})
quantization and metaplectic representation of ${\rm Sp}(1,\R)$, the map
on the
torus $\T^2$ defined by the standard unit symplectic matrix
\be
\left(
   \begin{array}{cc}
      0 & 1 \\
      -1 & 0
   \end{array}
\right)
\ee
This enables us to adapt to the elliptic map of the present case the
eigenvector
construction obtained in \cite{12} for the hyperbolic ones, based on the
determination of the suitable linear combinations of the the orthogonal
vectors (for fixed
$k\in\Z_{N}$)
\be
 \psi_{k,l}(q) = \frac1{\sqrt{p}}\exp\left [\frac{2\pi i}{p}(k{q}^2 +
lq)\right],
                  \quad k,l\in{\Z}_p
\ee
by action of the map
itself.\par\noindent
The orthonormality of the basis $\{\psi_{k,l}(q)\}: l=0,\ldots,p-1$, $k$
fixed requires $p$ prime and can be easily deduced using the well-known
properties of quadratic Gauss sums, in particular from the relation (see
e.g.\cite{8}, Chapter 9):
\be
\label{G}
\sum_{k=0}^{p-1}{\rm exp}\left[\frac{2\pi i}{p}(a{k}^2 + bk)\right]=
 \left\{ \begin{array}{ll}
   {\ds \epsilon_{p}p^{\frac{1}{2}}\left(\frac{a}{p}\right)
    \exp{\left[\frac{2\pi}{p}b^{2}{(4a)}^{-1}\right]} }
   & {\ds {\rm if}\quad a\not\equiv 0\;(\mod\; p)} \\
 {\ds p{\delta}_{b}^{0}} & {\rm if} \quad a\equiv 0 \;(\mod\; p)
 \end{array}
\right.
\ee
where
\be
\epsilon_{p} = \left\{ \begin{array}{ll}
                      1 &  p\equiv 1 \quad (\mod\; 4)\\
                      i &  p\equiv 3 \quad (\mod\; 4)
                          \end{array}
  \right.
\ee
Here and in what follows if $x\in\Z_p$ the symbol $x^{-1}$ denotes its
inverse in $\Z_p$, namely $x\cdot x^{-1}\equiv 1\;(\mod\,p)$. The
inverse is
unique because $\Z_p$ is a field since $p$ is prime.\par
 If $ p=4m + 1 $, $ -1$ is a quadratic residue of $p$ as we have already
recalled; then
we can denote
 $\lambda_{p}$ (or simply $\lambda$ where the context is clear) the
largest integer
($\mod\; p$) such that
\be
\lambda_{p}^{2}\equiv -1 \quad (\mod\; p)
\ee
and we denote $\Gamma$ a representative of the equivalence relation in
${\Z}_{p}^{\ast}$ :
\begin{equation}
 x\sim y \Longleftrightarrow y = {\lambda}^{s}x \quad{\rm for}\quad
s\in\{1,2,3,4\}
\end{equation}
 We can write
\be
\Z_{N}^{\ast}= \Gamma \cup (\lambda \Gamma)\cup (-\Gamma)
\cup (-\lambda \Gamma)
\ee
and we can choose $\Gamma$ in such a way that
\be
\Gamma \cup (\lambda \Gamma) = \{1,\ldots ,2m \}
\ee
Then we have:
\begin{proposition}
 A complete system of orthogonal eigenvectors of the operator ${\cal
F}_p$,  where
$p$ is prime such that $p=4m+1$, is given by:
\be
 \begin{array}{cl}
  {\psi}_{\{\overline{k},0\}} & {\rm with}\quad {\rm eigenvalue}\quad 1
\\
     \{\Phi_{j,r}:j\in\Gamma, r= 0,1,2,3\}&  {\rm with}\quad  {\rm
eigenvalue}\quad {i}^r
  \end{array}
\ee
where
\be
\Phi_{j,r}=
 \frac{1}{2}\sum_{s=0}^{3}{i}^{-sr}{\exp}\left[{\frac{\pi{i}}{p}\lambda
j^{2}
    \frac{1-{\lambda}^{rs}}{2}}\right]\;
\psi_{\overline{k},{\lambda}^{r}j},
\quad j\in\Gamma,\quad r=0,1,2,3
\ee
\end{proposition}
{\it Proof.}\par
 We have:
\begin{eqnarray*}
({\cal F} \psi_{\overline{k}, j})_{m}  = \frac{1}{\sqrt p}
\sum_{q=0}^{p-1}
                            \exp{\frac{2\pi i}{p} mq} \exp{\frac{2\pi
i}{N}(\overline k{q}^2 + jp)} =\\  =
\epsilon_{p}\left(\frac{\overline{k}}{p}\right) \exp\left [-\frac{2\pi
i}{p}{(m+j)}^2 + {(4\overline k)}^{-1}\right] = \\  =
\exp\left [\frac{2\pi
i}{p}(\overline k j^{2} +\overline k m^{2} + 2m\overline k j) \right] =
\\  =
\exp\left[\frac{\pi i}{p}\lambda j^{2}\right]
(\psi_{\overline k, \lambda j})_{m}
\end{eqnarray*}
 and, in general:
\be
{\cal F} \psi_{\overline{k}, \lambda^{s} j} =
         \exp\left[\frac{\pi i}{p}\lambda j^{2}\lambda^{2s}\right]
\psi_{\overline k,\lambda^{s+1} j}
\ee
where we have used the relation:
\begin{eqnarray*}
\left(\frac{\overline{k}}{p}\right)
    = \left(\frac{\lambda}{p}\right)\left(\frac{2^{-1}}{p}\right)
    = \left(\frac{\lambda}{p}\right)\left(\frac{2}{p}\right) = \\
    = (-1)^{(N-1)/4} (-1)^{(p^{2}-1)/8} = \left\{\begin{array}{cc}
           1 \cdot 1 & p \equiv 1 \;(\mod \;1) \\
           (-1) (-1) & p \equiv 3 \;(\mod\; 4) \end{array}
    \right.
\end{eqnarray*}

(See \cite{8}, Theorems 9.4, 9.5).

Then:
\begin{eqnarray*}
{\cal F} \Phi_{j,r}  = \frac{1}{2} \sum_{s=0}^{3} i^{-sr}
\exp\left[\frac{\pi
                 i}{p} \lambda j^{2} \frac{1-\lambda^{2s}}{2}\right]
                 {\cal F} \psi_{\overline k, \lambda^{s} j} = \\
                 = \frac{1}{2} \sum_{s=0}^{3} i^{-sr}
\exp\left[\frac{\pi
                 i}{p} \lambda j^{2} \frac{1-\lambda^{2s}}{2}\right]
                \exp\left[\frac{\pi i}{p} \lambda
j^{2}\lambda^{2s}\right] \psi_{\overline
                k,\lambda^{s}j} = \\
              = i^{r} \frac{1}{2} \sum_{s=0}^{3} i^{-(s+1)r}
\exp\left[\frac{\pi
                 i}{p} \lambda j^{2} \frac{1-\lambda^{2(s+1)}}{2}\right]
                \psi_{\overline k, \lambda^{s+1} j} = \\
             = i^{r} \Phi_{j,r}
\end{eqnarray*}
The orthonormality of the eigenvectors is implied by the orthonormality
of the
vectors
$ \psi_{\overline k, j}$ and a simple computation based on (\ref{G}).

It is now straightforward to obtain a complete system of eigenvectors of
the
sine Fourier transform operator $S = C^{p-1}\longrightarrow C^{p-1}$
whose
matrix elements are
\begin{equation}
(S)_{xy} = \frac{1}{\sqrt{p}} \sin(\frac{2\pi}{p}xy)\quad  x,y =
1,\ldots,p-1
\end{equation}
obtained by the discrete Fourier transform operator ${\cal F}_{p}$ :
\begin{equation}
S = \frac{{\cal F}_{p}-{\cal F}^{\ast}_{p}}{2i}
\end{equation}
Remark that a priori
$S$ is defined on ${\C}^{p}$. For the sake of simplicity we have
eliminated the
first row and the first column which are equal to zero and thus we
consider it
as an operator on ${\C}^{p-1}$. \par
We construct the eigenvalues of the operator $S$
by means of linear combinations of the real vectors :
\be
\begin{array}{l}
{\ds
\frac1{\sqrt{p}}\cos\left(\frac{2\pi}{p}ax^{2}\right)\sin\left(\frac{2\pi}{p}bx\right)}\\
{\ds
\frac1{\sqrt{p}}\sin\left(\frac{2\pi}{p}ax^{2}\right)\sin\left(\frac{2\pi}{p}cx\right)}
\end{array}
\ee
with a suitable choice of $ a, b, c.$  With the previous set of $\Gamma$
and
$\lambda$ we have (the proof is a straightforward verification based on
the
action of $\cal F$ specified in Proposition 3.1 and on (\ref{F1})):
\begin{proposition} Let $p$ be a prime such that $p = 4m +1$; then the
spectrum of the the operator $S$ consists in the eigenvalues
$-1,0,1$, and a set of corresponding eigenvectors is specified as
follows:
\be
\begin{array}{c} e_j(x) + e_{-j}(x),\quad j=1,\ldots,2m,\quad {\rm
eigenvalue}\quad  0 \\
  \Phi_{k,p}^{+}(x),\quad k\in\Gamma\quad {\rm eigenvalue}\quad {+1} \\
  \Phi_{h,p}^{-}(x)\quad h\in {\lambda}\Gamma\quad {\rm eigenvalue}\quad
{-1} \\
\end{array}
\ee
Here $e_{j}: j=1,\ldots,p-1$ is the canonical basis of ${\R}^{p-1}$,
$x=1,\ldots,p-1$ and
\begin{eqnarray*}
\Phi_{k,p}^{+}(x) =
                   \frac1{\sqrt{p}}\left[\cos\left( \frac{2\pi}{p}
\lambda{(2)}^{-1} x^{2}\right)\sin\left(\frac{2\pi}{p}kx\right)\right] +
\\
\frac1{\sqrt{p}}\left[\sin\left( \frac{2\pi}{p} \lambda(2)^{-1}(
k^{2} + x^{2}\right)\sin\left(\frac{2\pi}{p}\lambda kx\right)\right]
\end{eqnarray*}
\begin{eqnarray*}
\Phi_{h,p}^{-}(x) =
\frac1{\sqrt{p}}\left[\cos\left( \frac{2\pi}{p} \lambda{(2)}^{-1}
x^{2}\right)\sin\left(\frac{2\pi}{p}hx\right)\right] + \\
\frac1{\sqrt{p}}\left[\sin\left( \frac{2\pi}{p} \lambda(2)^{-1}( h^{2} +
x^{2}\right)\sin\left(\frac{2\pi}{p}\lambda hx\right)\right]
\end{eqnarray*}
\end{proposition}
\noindent
{\bf Remarks.}
\begin{enumerate}
\item
By standard estimates on Gauss sums (we omit the details) it is easily
seen that the
above eigenvectors are normalized as follows
\be
\|\Phi_{k,p}^{+}\|=\|\Phi_{h,p}^{-}\|=1+O\left(\frac{1}{\sqrt{p}}\right)
\ee
\item
By the same argument of Corollary 2.1, if $p=2N+1$ ($N=2m$) an
eigenvector basis
for $J$ is given by
\be
\label{base}
\begin{array}{c}
  \Phi_{k,2N+1}^{+}(x),\quad k\in\Gamma, \;x=1,\dots,N\quad {\rm
eigenvalue}\quad
{+1}
\\
  \Phi_{h,2N+1}^{-}(x)\quad h\in {\lambda}\Gamma,\;x=1,\dots,N\quad {\rm
eigenvalue}\quad {-1}
\\
\end{array}
\ee
\item
The choice of the index $h\in \lambda \Gamma$
labelling the vectors $\Phi_{h}^{-}$ is due to following property of the
eigenvector components:
\be
\Phi_{h}^{-} (x) = \Phi_{k}^{+} (\lambda x), \quad \quad {\rm if }\quad
\, h =
\lambda k
\ee
Different choices of index $h$ (always in a $\Gamma$-type subset of
$\Z_N$)
generate analogous relations among the eigenvector components.
\end{enumerate}
\vskip 0.2cm\noindent
Let us now go over to apply this construction to prove for that $p=2N+1$
prime of the form
$4m+1$ the matrix
$J$ does not admit any spin configuration among its eigenvectors. To see
this,
first remark that, by the same argument of Lemma 2.1 and Corollary 2.1,
the vector
${\ds
\chi_L\left(\frac{x}{p}\right): x=1,\ldots,N}$ is in the kernel of $S$.
We have
indeed
\par
\begin{eqnarray*}
\label{2.15}  (S\chi_L)_k=\frac1{\sqrt{p}}\sum_{x=1}^{N}{\rm
\sin}\left(\frac{2\pi}{p}kx\right)\chi_L(x)
=\frac1{2i\sqrt{p}}\left(\chi_L(k)-\chi_L(-k)\right)\sum_{x=1}^p\chi_L(x)
e^{\frac{2\pi i}{p}kx}\\
=\frac1{2i\sqrt{p}}\left(\chi_L(k)-\chi_L(-k)\right)i\sqrt{p} =0
 \end{eqnarray*}
since $\chi_L(-1)=1$ if $N=4m+1$ (see e.g.\cite{7}, Theorem 9.10).
\vskip 0.2cm
The second step is represented by the constatation that, when $p$ is a
prime of the
form $4m+1$, if a spin configuration is an eigenvector it cannot
distinguish
between the eigenvalue $1$ or $-1$ of $S$ (and hence of $J$), whose
eigenspaces
$V^+$ and $V^-$ have one and the same dimension as we have seen above.
This fact is
the key difference with $p$ prime of the form
$4m+3$: here the dimension of $V^+$ and
$V^-$ differs by one and the distinction is possible.
\begin{lemma} Let $p=2N+1, N=2m$, and denote once more $V^{\pm}$
the subspaces corresponding to the eigenvalues $\pm 1$ of $S$. \\
Then
there exists $v=\in V^+, v=(v_1,\ldots,v_p), v_k\in\{\pm 1\},
k=1,\ldots,p$ if and
only  there exists $u=\in V^+, u=(u_1,\ldots,u_p), u_k\in\{\pm
1\}k=1,\ldots,p$.
\end{lemma}
{\it Proof.} The vectors $\Phi_{k,p}^{+}(x)$ and $\Phi_{h,p}^{-}(x)$
defined
in (\ref{base}) span $V^+$ and $V^-$, respectively, and taken together
with the basis of $V^0$ form a
basis of $C^p$. Therefore if $v\in V^+$ there are coefficients $c_k$
such that
\[
v=\sum_{k\in\Gamma}c_k \Phi_{k,p}^{+}
\]
Now set
\[
u= \sum_{h\in\lambda\Gamma}d_h \Phi_{h,p}^{-}
\]
with $d_{\lambda k}=c_k$. The vector $u$ is obviously eigenvector of $S$
corresponding to the eigenvalue $-1$. Moreover, since ${\ds
\Phi_{h,p}^{-}(x)=\Phi_{k,p}^{+}(\lambda x)}$, we have
\[
u_x= \sum_{h\in\lambda\Gamma}d_h \Phi_{h,p}^{-}(x)=
\sum_{k\in\Gamma}c_k \Phi_{k,p}^{+}=v_{\lambda x}
\]
Therefore $u_x\in\{\pm 1\}\Longleftrightarrow v_{\lambda x}\in\{\pm 1\}
\Longleftrightarrow v_{x}\in\{\pm 1\}\;x=1,\ldots,p-1$ and this proves
the Lemma.
\vskip 0.2cm
Hence
\begin{proposition} If $p=2N+1, N=2m$ no antisymmetric eigenvector
(with eigenvalue $\pm 1$) of the matrix
$S$, and hence no eigenvector of $J$, can have all components $\pm 1$.
\end{proposition}
{\it Proof.} Consider the numbers ${\ds {\rm
\sin}\left(\frac{2\pi}{p}kx\right), k,x=1\ldots,p-1}$. Only $p-1=4m$ of
them are distinct. We can label them as ${\ds \mu_s={\rm
\sin}\left(\frac{2\pi}{p}s\right), s=1\ldots,\frac{p-1}{2}=4m}$. These
numbers
are all irrational (see e.g.\cite{6}, Thm 6.15). Now the eigenvector
relation
$S\chi_L =0$ yields, since the eigenvalue $0$ has multiplicity $2m+1$,
${\ds
p-2m-1=\frac{p-1}{2}=2m}$ independent relations with integer
coefficients among
the ${\ds 4m}$ numbers $\mu_s$. By the antisymmetry of $S$, these
conditions
are necessarily equivalent to the standard reflection conditions
\be
\label{AS}
\mu_s={\rm
\sin}\left(\frac{2\pi}{p}s\right)=-{\rm
\sin}\left(\frac{2\pi}{p}(p-s)\right)=\mu_{p-s},\quad s=1,\ldots,2m
\ee
 If there is an eigenvector $v=(v_1,\ldots,v_p)$ with
$v_s\in\{\pm 1\}\; ,s=1,\ldots,p-1$, the eigenvector condition $Sv=v$
yields
$p-1-m=3m$ independent conditions with integer coefficients because the
eigenvalue  $1$ has multiplicity $m$. Again, $2m$ of these conditions
are just
(equivalent to)  the conditions (\ref{AS}). We are thus left with $m$
independent relations with integer coefficients among the $2m$ numbers
$\mu_s$. However, by the former Lemma, the existence of $v$ as above is
equivalent to the existence of $u=(v_1,\ldots,u_p)$ with
$u_s\in\{\pm 1\}\; ,s=1,\ldots,p-1$ such that $Sv=-v$. Therefore, since
also the
multiplicity of the eigenvalue $-1$ is $m$, we get other $m$ independent
relations with integer coefficients among the numbers $\mu_s$. The $m$
relations $Sv=v$ are independent of the $m$ relations $Su=-u$ because
otherwise the vectors $u$ and $v$ would have non-zero components along
each
other, thus contradicting the orthogonality between $V^+$ and $V^-$. We
thus
end up with $2m$ linearly independent relations with integer
coefficients among
the $2m$ numbers $\mu_s$, and this contradicts their irrationality. This
proves
the statement as far as the matrix $S$ is concerned, and the assertion
for $J$
follows immediately by antisymmetry. This proves the proposition. \vskip
0.2cm
\noindent
{\bf Remark.} The above argument applies also to any $p=4m+1$ non prime
provided
\begin{itemize}
\item[(a)]
$V^+$ and $V^-$ have the same dimension and
\item[(b)]
Lemma 3.2 holds also for $p$ non prime.
\end{itemize}
Property (a) holds for $S$, and hence for $J$, because the eigenspaces
of
${\cal F}$ corresponding to the eigenvalues $i$ and $-i$ have one and
the
same dimension; we are however unable to prove property (b), even though
it
looks natural, because the eigenvector construction of Proposition 3.1
requires $p$ prime.
\vfill\eject

\end{document}